\documentclass[a4paper]{jpconf}
\usepackage{graphicx}
\usepackage{listings,multicol}
\usepackage{parcolumns}
\usepackage{xcolor}

\definecolor{dkgreen}{rgb}{0,0.6,0}
\definecolor{gray}{rgb}{0.5,0.5,0.5}
\definecolor{mauve}{rgb}{0.58,0,0.82}
\lstdefinestyle{myScalastyle}{
  frame=tb,
  float=*,
  language=scala,
  aboveskip=3mm,
  belowskip=3mm,
  showstringspaces=false,
  columns=flexible,
  basicstyle={\small\ttfamily},
  numbers=none,
  numberstyle=\tiny\color{gray},
  keywordstyle=\color{blue},
  commentstyle=\color{dkgreen},
  stringstyle=\color{mauve},
  frame=single,
  breaklines=true,
  breakatwhitespace=true,
  tabsize=3,
}

\begin{document}
\title{Using Big Data Technologies for HEP Analysis}

\author{Matteo Cremonesi$^2$, Claudio Bellini$^4$, Bianny Bian$^4$, Luca Canali$^1$, Vasileios Dimakopoulos$^1$, Peter Elmer$^5$, Ian Fisk$^3$, Maria Girone$^1$, Oliver Gutsche$^2$, Siew-Yan Hoh$^7$, Bo Jayatilaka$^2$,  Viktor Khristenko$^1$, Andrea Luiselli$^4$, Andrew Melo$^6$, Evangelos Evangelos$^1$, Dominick Olivito$^8$, Jacopo Pazzini$^7$, Jim Pivarski$^5$, Alexey Svyatkovskiy$^5$, Marco Zanetti$^7$}

\address{$^1$European Organization for Nuclear Research CERN, Geneva, Switzerland}
\address{$^2$Fermi National Accelerator Laboratory, Batavia, IL, USA}
\address{$^3$Flatiron Institute of the Simons Foundation, New York, NY, USA}
\address{$^4$Intel Corporation, Santa Clara, USA}
\address{$^5$Princeton University, Princeton, NJ, USA}
\address{$^6$Vanderbilt University, Nashville, TN, USA}
\address{$^7$University of Padova, Padova, Italy}
\address{$^8$University of California San Diego, La Jolla, USA}

\ead{matteoc@fnal.gov}

\begin{abstract}
The HEP community is approaching an era were the excellent performances of the particle accelerators in delivering collision at high rate will force the experiments to record a large amount of information. The growing size of the datasets could potentially become a limiting factor in the capability to produce scientific results timely and efficiently. Recently, new technologies and new approaches have been developed in industry to answer to the necessity to retrieve information as quickly as possible to analyze PB and EB datasets. Providing the scientists with these modern computing tools will lead to rethinking the principles of data analysis in HEP, making the overall scientific process faster and smoother.

In this paper, we are presenting the latest developments and the most recent results on the usage of Apache Spark for HEP analysis. The study aims at evaluating the efficiency of the application of the new tools both quantitatively, by measuring the performances, and qualitatively, focusing on the user experience. The first goal is achieved by developing a data reduction facility: working together with CERN Openlab and Intel, CMS replicates a real physics search using Spark-based technologies, with the ambition of reducing 1 PB of public data in 5 hours, collected by the CMS experiment, to 1 TB of data in a format suitable for physics analysis.

The second goal is achieved by implementing multiple physics use-cases in Apache Spark using as input preprocessed datasets derived from official CMS data and simulation. By performing different end-analyses up to the publication plots on different hardware, feasibility, usability and portability are compared to the ones of a traditional ROOT-based workflow.
\end{abstract}

\section{Introduction}

The scientific method is based on comparing predictions to experimental data, in order to confirm or disprove new theories. In high energy physics (HEP), such data are collected by an experimental apparatus that can detect fundamental particles once they are produced in the collision of beams provided by accelerators like the LHC at CERN.

Particle detection is an extremely complicated process. It consists in recording the physics quantities (like energy or flight path) of the particles generated in a collision. Such quantities are measured by the interaction of the particles with the different active components of the detector used to perform the experiment. At the LHC, given the complexity of the detector design, this process involves performing almost one hundred million independent measurements. All the measurements that happen in a single collision are collectively called an event.

The event is the unit which the whole HEP analysis process is based on. Event by event, detector signals (as well as simulated signals) must be converted into the physics properties associated to the particles that produced them. Complex algorithms are applied in order to reconstruct such information. This step is computationally expensive and it is usually organized centrally by each experiment, in order to make the best use of the available resources and to best serve the need of the researchers and the priorities of the experiment. The reconstructed events are provided to the collaboration of physicists in a shared format that is the input to the final analysis.

Usually the total size of the datasets as provided by central processing is too large to allow for interactive analysis. Researchers or groups of researchers exploring similar physics questions rely on several steps of data processing, filtering unnecessary events and eliminating unnecessary quantities from the original datasets  to get a manageable sub-sample. The optimization of this process is left to the individual.

In the next years the experiments at CERN will face a substantial increase in the data production rate due to a planned major upgrade of the accelerators. In order to ensure continuity in the production of high quality scientific results, the inefficiencies that are affecting the current analysis approach must be eliminated. The need to investigate alternative possibilities to perform physics analysis in a more efficient way is therefore becoming imperative.

Recently, new toolkits and systems have emerged outside of the HEP community to analyze Petabyte and Exabyte datasets in industry, collectively called "Big Data." These new technologies use different approaches and promise a fresh look at analysis of extremely large datasets. 

In this paper, we focus on the application of Apache Spark~\cite{Zaharia:2010:SCC:1863103.1863113} to the HEP analysis problem. We incorporate lessons learned from our previous investigations~\cite{DBLP:journals/corr/GutscheCEJKPSSS17} as well as new tools developed to enable HEP analysis in Apache Spark. Scalability and usability studies are performed and the latest findings are presented.

\section{The Traditional Analysis Workflow and Its Limitations}
\label{sec:root_workflow}

The traditional HEP analysis workflow is based on the usage of the ROOT framework~\cite{root}, a general, experiment-independent C++ toolkit. It provides statistical tools and a serialization format to persist reconstructed and transformed objects in files.

The centrally-produced datasets are provided in ROOT format, with a file-based data representation and a class structure with branches. The data management systems do not allow to extract branches efficiently from nested ROOT files, therefore physicists set up workflows that involve several steps of data processing, each one of them staging out intermediate outputs.

A first step of ntupling is performed in order to modify the information saved event by event. Immutable branches are duplicated in a disk-to-disk copy with the addition of new branches if needed, while unused ones are removed. At this stage, the information is selected to serve a smaller group of researchers performing similar physics measurements. Although the total size of the output is smaller, it is still too big to allow for interactive analysis. 

A second step that involves dropping uninteresting events (skimming) or additional unused branches (slimming) is therefore necessary to limit latencies. The output is a disk-to-disk copy where the immutable information is once again duplicated, but the class structure is translated into a "flat" format, in which events are rows of a table with primitive numbers or arrays of numbers as columns. At this stage, the information is usually selected to serve the scope of a single analysis. The size is reduced by an order of magnitude. 
As a last step, quantities from the final ntuples are aggregated and plotted as histograms.

These steps require the usage of grid and batch resources to exploit parallelization. Significant burden of tedious and time-consuming manual bookkeeping and failure re-submission is put on the individual analyst or analysis groups, resulting in an inefficient job splitting, with suboptimal parallelization. The worflow is convoluted and with limited interactivity. The analysis frameworks that support such workflows are group- or analysis-specific, often hardware-specific, limiting the portability and stimulating the multiplication of individual codes with similar functionalities. 
The duplication of immutable branches that happens at each stage of the workflow results in significant usage of storage space, making such an approach not sustainable on the long run.

\section{The Scalability Test}
\label{sec:spark_workflow}

In a previous usability study~\cite{DBLP:journals/corr/GutscheCEJKPSSS17} of Apache Spark, we implemented an analysis workflow by converting data into the AVRO~\cite{avro} format and uploaded it to the Hadoop file system (HDFS)~\cite{hdfs} of our development cluster. The biggest impediment to use the new technology as identified by the analysts was the need to convert the data in the new format. 

To  enable  Apache  Spark  to  understand  the  data  structures  of  the  ROOT  files  without the need to convert, a  library  called  spark-root~\cite{spark-root} was developed. It is based on a Java-only implementation of the ROOT I/O libraries which offers the ability to translate ROOT data structures into Spark DataFrames (DFs) and ResilientDistributedDatasets (RDDs). 

Another library, the Hadoop-XRootD connector~\cite{hadoop-xrootd-connector} was also developed in order to enable access to files stored in the EOS~\cite{eos} disk  storage  system  at CERN. The Hadoop-XRootD  connector is a Java  library  that  accesses  files  directly  through  the  XRootD  protocol~\cite{xrootd}  without  the  need  to import/export  data to  HDFS.

In  this  paragraph,  the latest results of the tests on  performance,  efficiency  and  scalability  of  these new  tools are presented. A  Spark workflow  that  reproduces  a  real  physics  analysis  was  executed  on  a Hadoop cluster  at  CERN.  This infrastructure is comprised of almost 1300 cores and 7TB of RAM. The input is data collected by the CMS~\cite{cms} experiment in 2011, publicly accessible and stored as ROOT files on EOS at CERN.

The tests were performed by executing the workflow for different size of the input data, in order to understand how the execution time scales with input size. As a second step, the same workflow was executed for a specific input size while scaling up the available resources (executors/cores and memory). 

The tests were repeated for two different instances of the EOS storage, namely EOS Public and EOS UAT, in order to identify if the network throughput and the storage infrastructure affect the performances. The EOS UAT instance consists of six servers, used exclusively for these tests.

The results of the first round of tests were obtained fixing to 32 MB the size of the ''readAhead'' buffer of the Hadoop-XRootD connector, which determines the amount of data that the connector will pre-fetch from the EOS storage service with every read call.

\begin{figure}
\caption{Performances for different input size, obtained with a ''readAhead'' buffer of 32 MB, 804 logical cores, and 2 logical cores per Spark executor.}
\label{fig:size_time_plot}
\center
\includegraphics[scale=0.25]{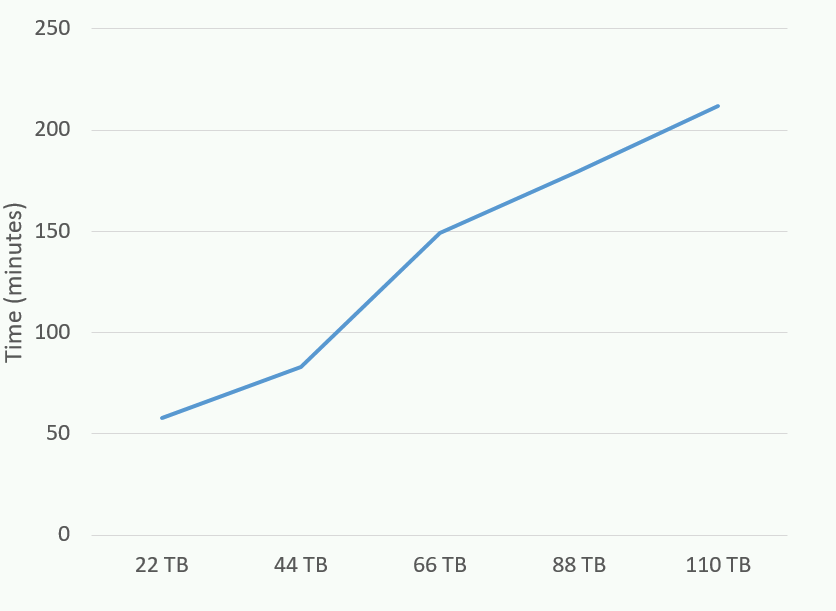}
\end{figure}

Table~\ref{table:size_time_plot} and Fig.~\ref{fig:size_time_plot} show a linear dependence between the input size and the execution time. The system is able to reduce 110 TBs in 212 minutes without any significant optimization.

\begin{table}
	\caption{The results of scalability tests for different input size, obtained with a ''readAhead'' buffer of 32 MB, 804 logical cores, and 2 logical cores per Spark executor.}
\label{table:size_time_plot}
	\centering
\begin{tabular}{cc}
 \hline
	Input Size                   & Execution Time                   \\ \hline \hline
	22 TB                             & 58 min                              \\ 
	44 TB                             & 83 min                              \\ 
	66 TB                              & 149 min                            \\ 
	88 TB                              & 180 min                            \\ 
	110 TB                            & 212 min                              \\ 
\hline
\end{tabular}
\end{table}

\begin{figure}
\caption{Performances for different number of cores, obtained with a ''readAhead'' buffer of 32 MB.}
\label{fig:cpu_time_plot}
\center
\includegraphics[scale=0.3]{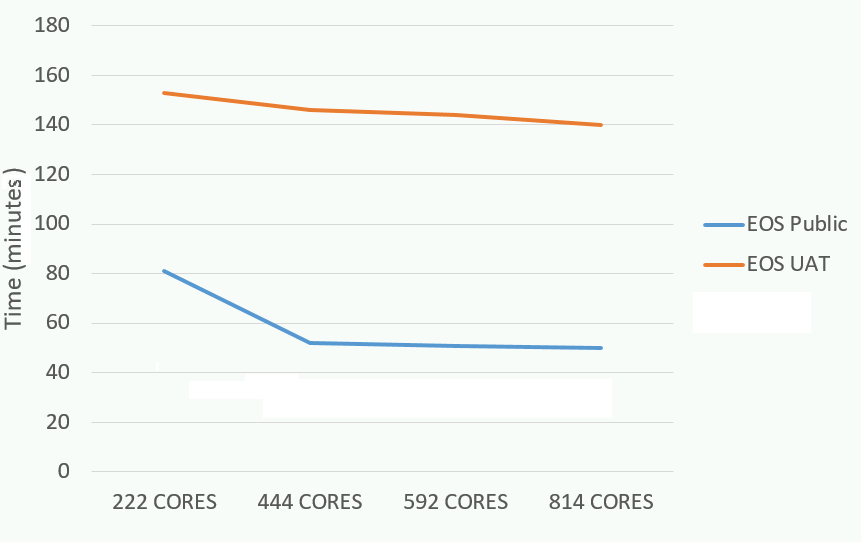}
\end{figure}

Table~\ref{table:cpu_time_res} and Fig.~\ref{fig:cpu_time_plot} show the results as a function of the allocated memory and cores, obtained by adjusting the number of Spark executors and fixing the memory per executor to 7GB. A plateau in the performance is reached at a specific memory value. This  effect  was  caused  by  the  saturation  of  the  available  network  bandwidth.  It  was  also evident  when  monitoring  the  total  throughput  of  the  network,  as  shown  in  Table~\ref{table:net_cores_res}.




\begin{table}
\caption{The results obtained scaling up the number of cores with 22 TB, with a ''readAhead'' buffer of 32 MB, for the different EOS instances used in this test.}
\label{table:cpu_time_res}
\centering
\begin{tabular}{cccc}
\hline
Number of Executors/Cores & EOS public & EOS UAT \\ \hline\hline

111/222                                                                                     & 81 min                                                                                            & 153 min                                                                                                                                \\ 

222/444                                                                                     & 52 min                                                                                            & 146 min                                                                                                                               \\ 

296/592                                                                                     & 51 min                                                                                            & 144 min                                                                                                                              \\ 

407/814                                                                                     & 50 min                                                                                            & 140 min                                                                                                                              \\ 
\hline
\end{tabular}
\end{table}

\begin{table}
\caption{Network throughput of EOS Public and EOS UAT access for different number of cores, obtained with a ''readAhead'' buffer of 32 MB.}
\label{table:net_cores_res}
\center
\begin{tabular}{ccc}
\hline
Cores & EOS public & EOS UAT     \\ \hline\hline

222                          & 15GBytes/s                        & 6GBytes/s   \\ 

444                          & 19GBytes/s                        & 7.5GBytes/s \\

592                          & 21GBytes/s                        & 7.5GBytes/s \\

814                          & 21GBytes/s                        & 7.5GBytes/s \\ 
\hline
\end{tabular}
\end{table}


Further investigations identified that the ''readAhead'' parameter of the Hadoop-XRootD connector would need further tuning. After multiple executions with different parameter sizes that varied between 32 MB and 16 KB, an optimal value of 64 KB was determined for this specific workload. This choice allowed the utilization of 8 logical cores per Spark executor (compared to the 2 cores per executor used in the previous test) due to the lower memory pressure. As a result, a dramatic increase in performance was achieved, in terms of execution time and of efficiency in the utilization of the underlying resources. Such improvement can be explained by considering the specific features of the ROOT file format, that allows for read operations that fetches only the data of interest, rather than scanning the entire input ROOT file. Therefore a relatively small value of the ''readAhead'' parameter optimizes the performances.

\begin{figure}
\caption{Performance of the tests for different input size in minutes, obtained with a ''readAhead'' buffer of 64 KB, 804 logical cores, and 8 logical cores per Spark executor.}
\label{fig:size_time_plot_sc-2}
\center
\includegraphics[scale=0.5]{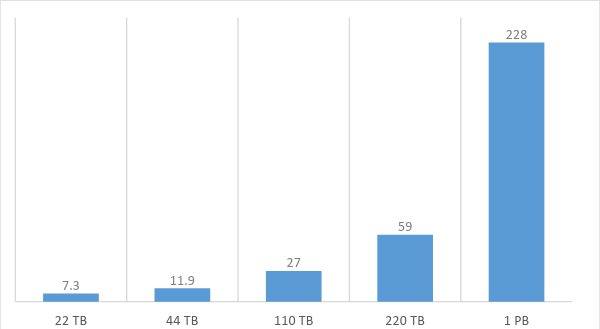}
\end{figure}

All the tests were re-executed with the new ''readAhead'' buffer size for the EOS public instance. The final results are summarized in Table~\ref{table:size-time-sc2} and Fig.~\ref{fig:size_time_plot_sc-2}.

\begin{table}
\caption{Aggregate results from re-executing the test by scaling up the input data size on EOS publicobtained with a ''readAhead'' buffer of 64 KB, 804 logical cores, and 8 logical cores per Spark executor.}
\label{table:size-time-sc2}
\centering
\begin{tabular}{ccc}
\hline

Input  & EOS  PUBLIC \\ \hline\hline

22 TB  & 7.3 mins     \\ 

44 TB  & 11.9 mins     \\

110 TB  & 27 mins     \\ 

220 TB  & 59 mins      \\

1 PB & 3.8 hours      \\ 
\hline
\end{tabular}
\end{table}

The new results prove the capability to reduce 1 PB of input data in less than four hours. 
Some key metrics of the workload were measured with a custom-developed Spark library and the results are shown in Table~\ref{table:metrics}. 

\begin{table}
\caption{Key workload metrics measured with Spark custom instrumentation for 1 PB of input, obtained with a ''readAhead'' buffer that varies between 16 KB and 64 KB, 804 logical cores, and 8 logical cores per Spark executor.}
\label{table:metrics}
\centering
\begin{tabular}{ccc}
\hline

Metric  & Total Time Spent \\ \hline\hline

Total execution time & 3000-3500 hours     \\ 

CPU time & 1200 hours    \\

Read time  & 1200 - 1800 hours, depending on the ''readAhead'' size    \\ 

Garbage collection time  & 200 hours      \\

\hline
\end{tabular}
\end{table}

The CPU time used by tasks for processing accounts for approximately 40\% of the execution time. The read time contributes to 40-50\% of the total, as expected considering that the data reside in an external service. Notably, garbage collection consumes only 7-8\% of the execution time.

\begin{figure}
\caption{Number of concurrent active tasks throughout the job execution for 1 PB of input, obtained with a ''readAhead'' buffer of 64 KB, 804 logical cores, and 8 logical cores per Spark executor.}
\label{paral}
\center
\includegraphics[scale=0.15]{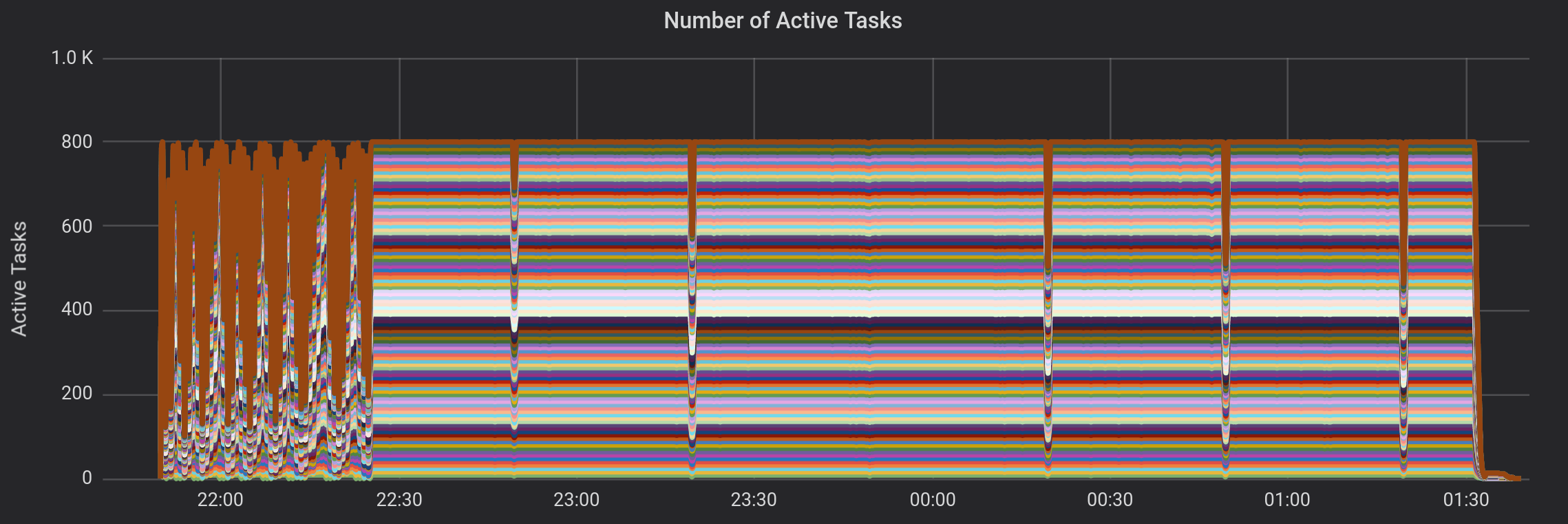}
\end{figure}

\begin{figure}
\caption{Executor CPU usage throughout the job execution for 1 PB of input, obtained with a ''readAhead'' buffer of 64 KB, 804 logical cores, and 8 logical cores per Spark executor.}
\label{cpuusage}
\center
\includegraphics[scale=0.15]{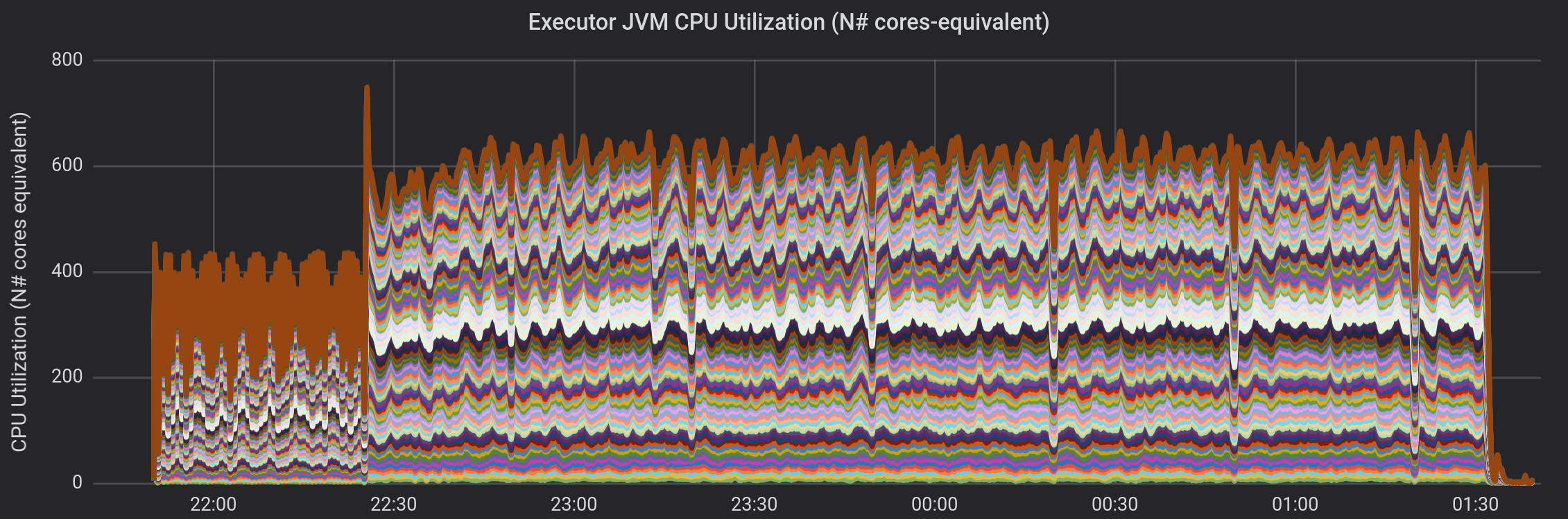}
\end{figure}

\begin{figure}
\caption{Read throughput from Hadoop-XRootD connector throughout the job execution for 1 PB of input, obtained with a ''readAhead'' buffer of 64 KB, 804 logical cores, and 8 logical cores per Spark executor.}
\label{through}
\center
\includegraphics[scale=0.15]{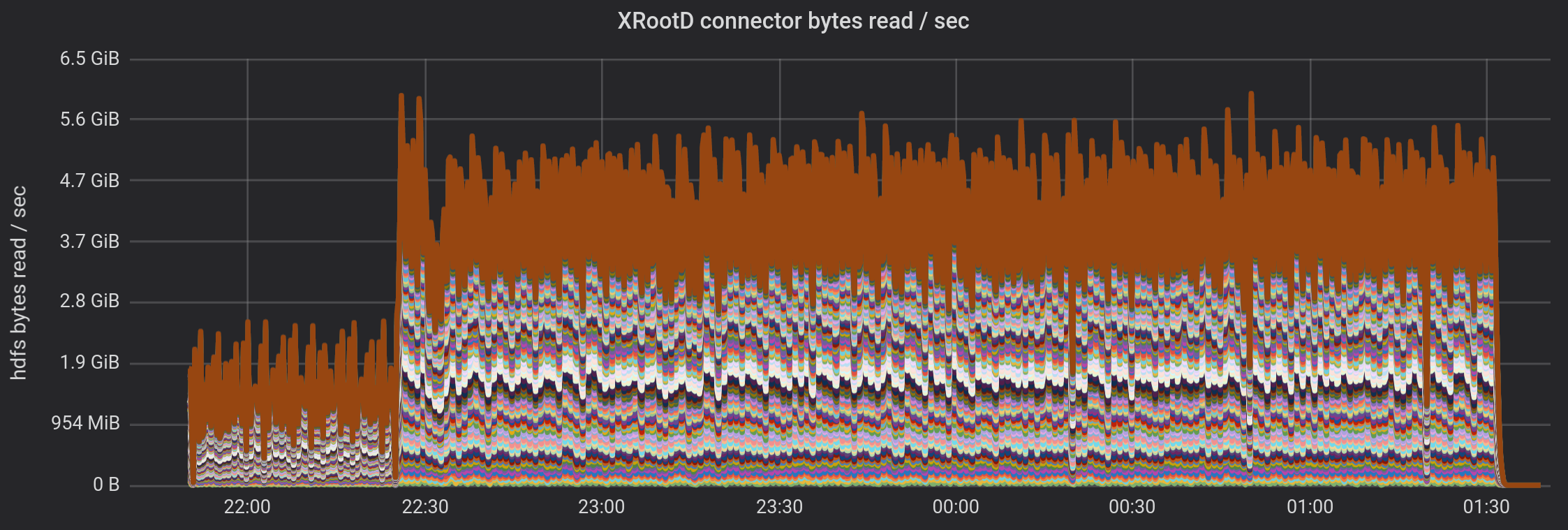}
\end{figure}

The Fig~\ref{paral} shows that the parallelization factor remained constant and close to optimal as the number of concurrent active tasks stayed at the maximum value (which equals the number of allocated cores for this job) for most of the job duration. This is a sign of both good scalability and proper usage of the underlying resources, as it is also demonstrated in Figs.~\ref{cpuusage} and~\ref{through}.

\section{The Usability Study}

The main goal of the usability test described in this section is testing the user experience, from the ability to setup and run a Spark-based analysis workflow to the portability of such workflow to different use cases and, most importantly, different clusters.

To perform this test, two similar workflows were tuned to run on different clusters. First, a Spark cluster at Vanderbilt University was used. It consists of 1000 cores with 5 TB of RAM. A second cluster hosted at the University of Padova, with 40 cores and 16 GB of RAM, was also employed in this test.
The two workflows shared a similar structure: load standard ROOT files as Spark DFs making use of the spark-root library, open them over XRootD with the Hadoop-XRootD connector, use Spark to transform DFs and aggregate them into histograms with the Histogrammar~\cite{jim_pivarski_2016_61418} package, produce plots and tables from the histograms.

The first step of the test consisted in verifying how easily such workflow can be set up by a newcomer. A first year undergraduate student in computer science, with no physics knowledge and limited computing knowledge, approached the issue. Starting from scratch, the student was able to learn the basic functionalities of the new tools and run the entire workflow in a day. The simplicity of the Spark workflow to be set up in a short timescale is a clear advantage when compared to the timescale that is needed for a newcomer once approaching the classical ROOT-based workflow.

The second step consisted in adapting to the Vanderbilt cluster the workflow tuned to run at the Padova cluster. The major showstopper encountered was the environment setup. This could be solved by for example developing a shared library that generalizes the site configuration. Additional improvement is also required for the packaging of the Hadoop-XRootD connector in order to make the tool more automatically deployable, avoiding manual configuration. A possible solution that has been explored would be deploying the connector through a  management framework for resource sharing as Apache Mesos~\cite{mesos}. Both these developments are currently a work in progress.

\section{Conclusions}

We presented studies of executing the traditional HEP analysis workflow on Apache Spark. The efficiency of the application of the new tools has been evaluated both quantitatively, by measuring the performances, and qualitatively, focusing on the user experience. Our studies identified some bottlenecks which we managed to overpass and underlined the need to scale up the Spark infrastructure and to generalize the site configuration. The scaling behavior results are promising and overperformed the original goal of reducing 1 PB to 5 hours, accomplishing the task in less than 4 hours.

\section{Acknowledgments}

We would like to thank the CMS collaboration and the LHC to provide the data for the use case and the ROOT based workflow. This work was partially supported by Fermilab operated by Fermi Research Alliance, LLC under Contract No. DE-AC02-07CH11359 with the United States Department of Energy, and by the National Science Foundation under grants ACI-1450377 and PHY-1806612, and Cooperative Agreement PHY-1120138.

\section*{References}
\bibliography{main}

\providecommand{\newblock}{}
\begin{thebibliography}{10}
\expandafter\ifx\csname url\endcsname\relax
  \def\url#1{{\tt #1}}\fi
\expandafter\ifx\csname urlprefix\endcsname\relax\def\urlprefix{URL }\fi
\providecommand{\eprint}[2][]{\url{#2}}

\bibitem{Zaharia:2010:SCC:1863103.1863113}
Zaharia M, Chowdhury M, Franklin M~J, Shenker S and Stoica I 2010 {\em
  Proceedings of the 2Nd USENIX Conference on Hot Topics in Cloud Computing\/}
  HotCloud'10 (Berkeley, CA, USA: USENIX Association) pp 10--10
  \urlprefix\url{http://dl.acm.org/citation.cfm?id=1863103.1863113}


\bibitem{DBLP:journals/corr/GutscheCEJKPSSS17}
Gutsche O, Cremonesi M, Elmer P, Jayatilaka B, Kowalkowski J, Pivarski J,
  Sehrish S, Surez C~M, Svyatkovskiy A and Tran N 2017 {\em CoRR\/} {\bf
  abs/1703.04171} \urlprefix\url{http://arxiv.org/abs/1703.04171}

\bibitem{root}
Brun R and Rademakers F 1997 {\em Nuclear Instruments and Methods in Physics
  Research Section A\/} {\bf 389} 81 -- 86 ISSN 0168-9002 new Computing
  Techniques in Physics Research V
  \urlprefix\url{http://www.sciencedirect.com/science/article/pii/S016890029700048X}

\bibitem{avro}
Apache avro \urlprefix\url{http://avro.apache.org}

\bibitem{hdfs}
Shvachko K, Kuang H, Radia S and Chansler R 2010 {\em Proceedings of the 2010
  IEEE 26th Symposium on Mass Storage Systems and Technologies (MSST)\/} MSST
  '10 (Washington, DC, USA: IEEE Computer Society) pp 1--10 ISBN
  978-1-4244-7152-2 \urlprefix\url{http://dx.doi.org/10.1109/MSST.2010.5496972}

\bibitem{spark-root}
Khristenko V and Pivarski J 2017 diana-hep/spark-root: v0.1.14\_pre1 release
  \urlprefix\url{https://doi.org/10.5281/zenodo.1019880}

\bibitem{hadoop-xrootd-connector}
Motesnitsalis V hadoop-xrootd-connector
  \urlprefix\url{https://gitlab.cern.ch/awg/hadoop-xrootd-connector}

\bibitem{eos}
Eos: Large disk storage at cern \urlprefix\url{https://eos.web.cern.ch}

\bibitem{xrootd}
Dorigo A, Elmer P, Furano F and Hanushevsky A 2005 Xrootd- a highly scalable
  architecture for data access WSEAS Transactions on Computers

\bibitem{cms}
Chatrchyan S~e~a (CMS Collaboration) 2008 {\em JINST\/} {\bf 3} S08004

\bibitem{jim_pivarski_2016_61418}
Pivarski J, Svyatkovskiy A, Schenck F and Engels B 2016 histogrammar-python:
  1.0.0 \urlprefix\url{https://doi.org/10.5281/zenodo.61418}

\bibitem{mesos}
Hindman B, Konwinski A, Zaharia M, Ghodsi A, Joseph A, Katz R, Shenker S and Stoica I 2011  {\em Mesos: A Platform for Fine-Grained Resource Sharing in the Data Center\/}
NSDI. {\bf 11}: 22-22.



\end{thebibliography}

\end{document}